\documentclass[journal]{IEEEtran}
\usepackage{latexsym,,amssymb,amsmath,graphicx,epsf,cite,bbm,float}
\usepackage{ifpdf}
\usepackage{epstopdf,mathtools}

\usepackage{algorithm,algorithmic}
\usepackage{amsmath,amssymb,bm}
\usepackage{amsfonts,dsfont,color,bbm,subcaption}

\def\ninept{\def\baselinestretch{1}}
\ninept

\newcommand{\abs}[1]{|#1|}

\usepackage{hyperref,amsthm}

\newtheorem{theorem}{Theorem}

\newtheorem{lemma}[]{Lemma}

\newtheorem{proposition}[]{Proposition}

\newtheorem{remark}[]{Remark}

\newtheorem{assumption}[]{Assumption}

\newtheorem{example}[]{Example}

\newtheorem{notion}[]{Notion}

\begin{document}

\title{Merging Knockout and Round-Robin Tournaments: A Flexible Linear Elimination Tournament Design} 
\author{\IEEEauthorblockN{Kaan Gokcesu}, \IEEEauthorblockN{Hakan Gokcesu} }
\maketitle

\begin{abstract}
We propose a new tournament structure that combines the popular knockout tournaments and the round-robin tournaments. As opposed to the extremes of divisive elimination and no elimination, our tournament aims to eliminate the participants as linearly as possible as a form of subtractive elimination. Our design is flexible in the sense that it can be adapted to any number of players $N$ and any number of matches $M$. Our design satisfies many properties that are desirable for a tournament to select a winner and can be adapted to rank all the participating players.
\end{abstract}

\section{Introduction}
We study the design of tournaments, which is scheduling the pairwise comparisons between participating competitors (players) \cite{sziklai2021efficacy}. There are varying formulations for such designs. While, in the literatures of sports, economics and management sciences, the formulation includes the distribution of some effort (intensity) for the sake of achieving some goal \cite{lazear1981rank,rosen1986prizes,taylor1995digging,prendergast1999provision,szymanski2003economic,orrison2004multiperson,brown2014selecting,bimpikis2019designing}; we consider the effort to be constant across the competition \cite{sziklai2021efficacy}. This formulation assumes that the competitors perform at their peak in all matches, which manifests itself in a lot of situations such as high stakes competitions, e.g., elections \cite{klumpp2006primaries}, musical competitions \cite{ginsburgh2003expert}, contests for crowdsourcing \cite{hou2021optimal}, sports tournaments \cite{palacios2009field}, innovation contests \cite{harris1987racing,yucesan2013efficient,ales2017optimal}.

There exist two traditional tournament formats: knockout tournaments and group tournaments. 
\begin{itemize}
	\item In the knockout tournaments, the most fundamental one is the single elimination tournament. Here, after a round of match-ups between the contestants, the losers are eliminated while the winners continue to compete. This tournament style has been extensively studied, especially in the fields of economics and statistics \cite{hartigan1968inference,israel1981stronger,hwang1982new,horen1985comparing,knuth1987random,chen1988stronger,edwards1998non,schwenk2000correct,glickman2008bayesian,vu2011fair,groh2012optimal,prince2013designing,krakel2014optimal,hennessy2016bayesian,karpov2016new,adler2017random,dagaev2018competitive,karpov2018generalized,arlegi2020fair,kulhanek2020surprises,arlegi2022can}. 
	\item In the group tournaments, the most fundamental one is the round-robin tournament. Here, all players compete against all other players and they are ranked based on their cumulative match results according to some rules \cite{harary1966theory,rubinstein1980ranking}.
\end{itemize}

There also exist multi stage tournaments which are combinations of knockout and group tournaments. The tournament design problem is in a way a sequential learning problem \cite{gokcesu2017online,gokcesu2018sequential}, where sequential matches determine the rankings.

When choosing a tournament format, various factors come into play. The most prominent ones are:
\begin{enumerate}
	\item fairness
	\cite{cea2020analytics,goossens2012soccer,guyon2015rethinking,guyon2018fairer,guyon2020risk,kendall2010scheduling,laliena2019fair,van2020handling,wright2014or}
	\item promoting competitiveness \cite{chater2021fixing}
	\item incentive compatibility \cite{csato2020incentive,csato2021tournament,csato2022quantifying,dagaev2018winning,pauly2014can,preston2003cheating,vong2017strategic}, 
	\item maximizing attendance \cite{krumer2020testing}
	\item minimizing rest mismatches \cite{atan2018minimization}
	\item revenue allocation \cite{bergantinos2020sharing,petroczy2021revenue}.
\end{enumerate}
 
The tournament design problem has a very complex structure with multitude of factors in consideration. Nonetheless, most of these factors have a common denominator, which is the fair ranking of the players. Strong contending players should continue to play and weak inconsequential players should not. For example, in multi stage tournaments, the players should be allocated (set partition \cite{gokcesu2021efficient,gokcesu2021quadratic,gokcesu2022linearithmic}) to groups in a fair manner. Moreover, a tournament's efficiency, i.e., how well it can rank the participating players according to their strength, is very important (if not the most). Note that the player strengths are hidden and the tournament needs to create this ranking based on the pair-wise match results as well as possible. Therefore, one can formulate tournaments as online learning problems \cite{gokcesu2017online,gokcesu2018sequential}, since sequential matches decide the relevant rankings. The match results are noisy in the sense that the stronger player may not always be the winner (the underdog can win). However, the winning probability of the stronger player monotonically increases \cite{gokcesu2021optimally,gokcesu2021anytime} with the strength difference between the matched-up players. Henceforth, the question becomes how to design the relevant loss functions \cite{gokcesu2021generalized,gokcesu2022nonconvex}. Such a loss function will probably have a very convoluted structure and potentially non-convex (hard to solve) \cite{gokcesu2021regret,gokcesu2022low}. Since it has a very convoluted structure, nonparametric analysis \cite{gokcesu2022natural,gokcesu2021nonparametric} is desirable. Because of these complexities, tournament design is a relatively hard problem.

If we approach this problem from a probabilistic perspective, it is possible to derive the probabilities of each player finishing in each rank for a small set of players based on some match result heuristics \cite{david1959tournaments,glenn1960comparison,searls1963probability}. However, this becomes increasingly more difficult as the number of players and the number of matches increase. If the tournament also has a convoluted structure, this method becomes intractable \cite{sziklai2021efficacy}.
For this reason, the standard approach is using Monte Carlo simulations. The work in \cite{appleton1995may} 
determines the likelihood of the best player winning the tournament in a variety of competitions.
The work in \cite{mcgarry1997efficacy} shows the efficiencies in ranking for traditional tournaments for eight players in a variety of initial conditions.
The work in \cite{mendoncca2000comparing} study varying ranking techniques for round-robin tournaments in their efficiency to create rankings equivalent to the strength of the players. The work in
\cite{marchand2002comparison} calculates the likelihood of a player in the top seed winning in standard and random knockout tournaments and determines that, unexpectedly, their outcomes are not much different from each other. The authors in \cite{ryvkin2008predictive} proclaim that the efficiency of determining the strongest player in the knockout and round-robin tournaments is non-monotonic with the number of players for heavy-tailed strength distributions among the players. In \cite{ryvkin2010selection}, the author studies alternative measures of selection efficiency.

Sometimes, we may not need the whole ranking and what is most important is who becomes the champion. Here, the tournament efficiency is determined by how likely the strongest player wins the tournament, which is especially true for sports tournaments. However, even such a simpler efficiency metric requires extensive assumptions to simulate and analyze. The work in \cite{scarf2009numerical} gives a numerical study of the sports tournament designs, including a comprehensive list of formats in practice, a simulation framework and their evaluation metrics.
The work in \cite{scarf2011numerical} continues this numerical analysis and studies the influence of the seeding policy on the tournament outcomes for the World Cup Finals. The work in \cite{goossens2012comparing} compares different league formats considered by the Belgian Football Association with respect to the match importance.
The work in \cite{annis2006comparison} compares potential playoff systems for NCAA IA football, where the systems differ in their number of playoff teams,  their selection and seeding. The work in \cite{lasek2018efficacy} studies the efficiency of league formats in their ability to rank teams in European association football competitions.
The work in \cite{csato2021simulation} compares different tournament designs for the World Men’s Handball Championships.

We may also have some past performances as an initial estimate for the rankings and the tournament can adopt this as a seeding rule. A better seeding of the players in the tournament may improve its efficiency, however, it is not easy to do so. After all, if we have a really good estimate for the power rankings, we may not even need a tournament in the first place. Interestingly, real data suggests that the performance in tournaments can be very different from any existing past performance; which makes seeding a somewhat marginal component in improving the efficiency of the tournaments \cite{sziklai2021efficacy}. This claim is supported by the study in the UEFA Europa League and the UEFA Champions League \cite{engist2021effect}, which shows that the seeding by itself is insufficient in its contribution to the success of better teams.

Nevertheless, in all analyses, more matches result in better estimations of the power rankings \cite{lasek2018efficacy,csato2021simulation}, which is statistically intuitive.
However, it becomes a concern especially if the number of participating players are too large. Furthermore, more general, flexible tournament designs are needed since the number of participants may substantially differ across competitions. The works in \cite{appleton1995may,mcgarry1997efficacy} have studied tournament designs with $8$ or $16$ players. The works in \cite{sziklai2021efficacy,scarf2009numerical} studies the tournaments with $32$ competitors. Although they study various tournament structures, the limitations on the number of participants is problematic. 

All in all, we tackle the tournament design problem from the perspective of how to best design a tournament given a limited number of matches. In this perspective, the knockout tournaments and round-robin tournaments are the two ends of the spectrum. It seems redundant to match up the same players against each other, hence, it is better to design different, more informative, match ups as much as possible \cite{sziklai2021efficacy}. However, we also cannot rank the players without at least $\log_2 N$ matches since the match results are binary and the rank can be represented as an at least $\log_2 N$ length binary sequence. To this end, we aim to naturally combine the knockout and group settings to create a flexible near linear elimination tournament.

\section{Linear Elimination Tournament Design}
In a seeded tournament let $n\in\{1,2,\ldots,N\}$, where $N$ is number of participants, be the indices of different players (or teams).

\begin{assumption}
	The player identified by $n$ is ranked $r_n=n$ at the beginning. The players are seeded such that the highest ranked player is $n=1$ and the lowest ranked player is $n=N$ based on some metric (such as the past performance).
\end{assumption}

In each round of the tournament, we have the following successive stages:
\begin{enumerate}
	\item Determination of match-ups
	\begin{itemize}
		\item At start of every round, match-ups between players are decided. 
		\begin{assumption}
			$N$ is even to ensure every player participates in a match, i.e., to minimize rest mismatches.
		\end{assumption}
		\begin{assumption}
			We disregard any unfair bias such as home court advantage.
		\end{assumption}
		
		\item After the matches, we have the results of which player won and lost. 
		
		\begin{assumption}
			We do not consider ties in the matches, i.e., there should always be a winner and a loser. 
		\end{assumption}
	\end{itemize}
	\item Re-ranking of the players
	\begin{itemize}
		\item Using these results, the players are re-ranked accordingly. 
	\end{itemize}
	\item Elimination of players
	\begin{itemize}
		\item After the new rankings, some players are eliminated from the tournament. 
		\begin{assumption}
			Even number of players are eliminated to ensure that every player can be matched-up in the successive rounds.
		\end{assumption}
	\end{itemize}
\end{enumerate}

\begin{remark}
	In both knockout and round-robin tournaments a similar fashion is followed in general. After the last round the last standing player or the first ranked player is crowned the champion.
\end{remark}

If all players that lose each round are eliminated, the tournament can end at best in $\log_2 N$ rounds. The question is how the tournament should be designed given the number of contestants $N$ and the number of matches $M$.

While designing such a tournament, we have to take note of few important factors as described in the introduction.
\begin{itemize}
	\item We need to ensure that both the re-ranking and the elimination stages are fair.
	\item For incentive compatibility and to promote competitiveness, we need to reward the winners and penalize the losers.
	\item To maximize attendance, we need to create high stakes matches as much as possible. Moreover, the tournament should gradually match up stronger players with each other.
	\item For tournament efficiency, weaker players need to be eliminated and stronger players should continue the tournament.
\end{itemize}

\subsection{Determination of Match-ups}
For the determination of the match-ups, we have the following notion.
\begin{notion}
	Stronger players should have higher chances of winning the tournament.
\end{notion}

This notion is traditionally followed in all kinds of tournament or competition designs, where the goal is to maximize the chances of the strongest player winning the tournament. To this end, in our design, the tournament follows a snake matching system in the sense that the first ranked plays the last ranked and the second ranked plays the second to last ranked etc., i.e., if players $i$ and $j$ match up, we have
\begin{align}
	r_i+r_j=N+1.
\end{align}

Since the rankings and the total number of remaining players change over time, we represent them as time dependent, i.e., after any $t^{th}$ match-ups, we have
\begin{align}
	N_t,& &&\forall t,\\
	r_{t,n},& &&\forall n,t,
\end{align} 
and if $i$ and $j$ match-up at $t^{th}$ round, we have
\begin{align}
	r_{t-1,i}+r_{t-1,j}=N_{t-1}+1.
\end{align}

Note that, before any match-up, i.e., $t=0$, we have $N_0=N$ and $r_{0,n}=n$. 
 
The tournament ends when $N_{\tau}=2$ for some $\tau$ and the winner is crowned the champion.

\subsection{Fair Re-ranking}

Note that, at each round $t$, the given rankings are our best estimates so far. To determine the re-ranking rules, we first make the following notion.
\begin{notion}\label{no:nonincdec}
	The ranking of a loser and a winner after a match-up should be nonincreasing and nondecreasing respectively. 
\end{notion}

This notion is observed in all kinds of tournament designs, where the winners are rewarded and the losers are penalized. Moreover, our only observations in each round are the results of the determined matches. To this end, we make the following notion.

\begin{notion}\label{no:order}
	The ranks of losers and winners after a match-up round should be ordered as before the match-up round individually.
\end{notion}

Moreover, the change in rankings, i.e., rank increase for the winners and the rank decrease for the losers should not be biased, hence, fair. To ensure this, we make the following notion.

\begin{notion}\label{no:together}
	If two consecutively ranked players both won (or lost), their increase (or decrease) in the rankings should be equal.
\end{notion}  

\subsection{Formulating Re-ranking as an Optimization Problem}
We can think of the re-ranking as an optimization problem, where we want to re-rank the players as much as possible whilst conforming to the notions. To construct the optimization problem, we need to define a gain or loss function. 

Let us have a binary vector of results $\boldsymbol{b}=\{b_n\}_{n=1}^N$, where $b_n=1$ if $n^{th}$ player won or $b_n=0$ if that player lost. Observe that we have $b_n\in\{0,1\}, \forall n$ and $b_n+b_{N+1-n}=1$ because of the match-ups. 

\begin{remark}
	Note that if the players were perfectly ranked and the matches resulted according to the expectations, we would have $b_n=1, n\in\{1,2,\ldots,N/2\}$ and $b_n=0, n\in\{N/2+1,N/2+2,\ldots,N\}$.
\end{remark}

Let $\boldsymbol{r}=\{r_n\}_{n=1}^N$ be the new ranks of $n$-ranked players. Since we want to re-rank them as much as possible with the new information (the match results), we can define the objective as
\begin{align}
	\max_{\boldsymbol{r}}\sum_{n=1}^N\abs{r_n-n},\label{eq:obj}
\end{align} 
given $\boldsymbol{b}$. However, this optimization problem as itself exploits the results of the matches, which can be observed in the following. 

\begin{proposition}
	The global maximizer of the objective function in \eqref{eq:obj} (that conforms to the notions given $\boldsymbol{b}$) is given by the new ranks $r_n$, which are the highest ranked winner to the lowest ranked winner, then the highest ranked loser to the lowest ranked loser.
	\begin{proof}
		First of all, this solution conforms to the notions. Moreover, the new ranks for the winners and the losers are upper and lower bounds respectively from \autoref{no:order}. Hence, this ranking is a global maximizer from \autoref{no:nonincdec}, which concludes the proof.
	\end{proof}
\end{proposition}

However, such a ranking mostly disregards the initial ranking. Since the initial rankings give us some form of estimate for the true power rankings, we should not disregard it. We observe that the same global maximizer solution is a global minimizer for the following optimization problem. We can consider the number of changes in the vector $\boldsymbol{b}$ as a loss (a measure of how well we predicted the rankings), i.e., 
\begin{align}
	\min_{\boldsymbol{r}}\sum_{n=1}^{N+1}\abs{\tilde{b}_n-\tilde{b}_{n-1}},\label{eq:objB}
\end{align} 
where $\tilde{b}_{r_n}=b_n$ for a given $\boldsymbol{b}$; $\tilde{b}_{0}=1$, $\tilde{b}_{N+1}=0$ are dummy results. The global maximizer of \eqref{eq:obj} is a global minimizer of \eqref{eq:objB}, where the minimum value is $1$. We observe that to increase the objective in \eqref{eq:obj}, while abiding by the notions, we need to simultaneously decrease the objective in \eqref{eq:objB}. To guarantee exploration whilst utilizing the match results (exploitation), we can put a constrained on the decrease on \eqref{eq:objB}.
To this end, to limit the change in the rankings between our old and new estimates, we regularize the reranking objective function such that we want to maximize one of them whilst constrained by the other. 

Although limiting the number of changes in the new ranks and minimizing the path change of $\boldsymbol{b}$ can also be considered, putting a constraint on the path change of $\boldsymbol{b}$ and maximizing the new rank changes is more natural since it is the initial objective. Moreover, we observe that while abiding by the notions, the value of the objective in \eqref{eq:objB} can only be odd. Thus, one such constraint is limiting the path change of $\tilde{\boldsymbol{b}}$ two less than the original path change of ${\boldsymbol{b}}$. 

Hence, unless the path change of $\boldsymbol{b}$ is $1$, the optimization function is .
\begin{align}
	\max_{\boldsymbol{r}}\sum_{n=1}^N\abs{r_n-n}:&&\sum_{n=1}^{N+1}\abs{\tilde{b}_n-\tilde{b}_{n-1}}=\sum_{n=1}^{N+1}\abs{b_n-b_{n-1}}-2.
\end{align}   
where $\tilde{b}_{r_n}=b_n$ and $\tilde{b}_{0}=1$, $\tilde{b}_{N+1}=0$, ${b}_{0}=1$, ${b}_{N+1}=0$ are dummy results.

\begin{theorem}
	After a match-up, switching the places of consecutive losers with winners is a solution to the optimization function.
	\begin{proof}
		To satisfy \autoref{no:together}, we observe that the consecutive ranked teams with same match results cannot be separated. Hence, we can only move them together in the rankings. To satisfy \autoref{no:nonincdec}, winners either stay in the same ranks or move up (the converse for the losers). Thus, our only move is swapping the places of consecutive winners with immediate higher ranked losers. Just one swap between any such groups will decrease the path change by $2$. However, this may not be the optimal solution. We observe that simultaneous swapping of all such groups also decreases the path change by $2$. Thus, we swap all such groups with each other to maximize the total number of rank changes, which concludes the proof.
	\end{proof} 
\end{theorem}

This re-ranking is intuitive in the sense that each player rises or falls in the rankings by utilizing exclusively the limited information about the match results.

\begin{remark}
	After a match up between players $i$ and $j$, the winner climbs up the rankings as much as the loser climbs down. Hence, each match-up satisfies a zero-sum game in the sense that they are equally rewarded or penalized in their cumulative rank changes.
\end{remark}

\begin{example}
	For example, let the players be ranked $$\boldsymbol{r}^{old}=\{1,2,3,4,5,6,7,8,9,10,11,12,13,14\}$$ at some round $t$ for $N=14$. Let these players have the match-up results $$W,L,L,W,W,W,L,W,L,L,L,W,W,L$$
	where the binary values $1$ and $0$ is changed with $W$ (win) and $L$ (lose) for ease of understanding. Note that this sequence is odd symmetric around its middle because of the match-up rule. Our re-ranking results in the following new rankings $$\boldsymbol{r}^{new}=\{1,4,5,6,2,3,8,7,12,13,9,10,11,14\}.$$
\end{example}

\begin{remark}
	For higher exploitation, the re-ranking can be considered in a modular manner and applied multiple times, where we decrease the value of the objective in \eqref{eq:objB} by $2$ at each time.
\end{remark}

\subsection{Elimination of Players}

For the player eliminations, to avoid resting periods for players, we have the following notion.

\begin{notion}
	Every player should match-up every round.
\end{notion}

To this end, to preserve valid match-ups for all players, the eliminated players should be even numbered, i.e.,
\begin{align}
	N_{t-1}-N_t=0\pmod{2}.
\end{align}

Moreover, we make the following notion to keep the match-ups and the tournament meaningful.
\begin{notion}
	Every player still in the running should have a chance to win it all.
\end{notion}

For this reason, we first postulate that the eliminated players can only be from the set of players who lost in their last round. For the elimination part, we eliminate the lowest ranking set of players that lost their last match. 

Given $N$ players and $M$ competition rounds, we determine the eliminated number of players as follows.
Since we somehow combine knockout and round-robin tournaments, we distribute the eliminations approximately linearly. In knockout tournaments, the players are eliminated in a multiplicative fashion, while in the round-robin no players are eliminated. Therefore, the linear elimination is somewhere in between. Since we only eliminate the losers, we need to satisfy
\begin{align}
	N_{t}\leq 2N_{t+1}, &&\forall t.
\end{align}

Thus, we linearly distribute the eliminated players across different rounds as much as possible whilst satisfying the constraints.
Hence, we formulate the following objective
\begin{align}
	\min_{\boldsymbol{N}}\sum_{t}\abs{(N_{t}-N_{t+1})-(N_{t-1}-N_{t})}: && 	N_{t}\leq 2N_{t+1}, \forall t, \label{eq:N}
\end{align}

\begin{lemma}\label{thm:noninc}
	The resulting sequence $\boldsymbol{N^*}$ has nonincreasing change, i.e.,
	\begin{align}
		N_{t}-N_{t+1}\leq N_{t-1}-N_{t}, &&\forall t.
	\end{align}
	\begin{proof}
		Let us assume it is not nonincreasing. In such a scenario, if we move the excess elimination that does not satisfy nonincreasing behavior to the first round for all such places, we will have a smaller path change in the number of eliminated players. Hence, in the optimal solution, we need to have a nonincreasing elimination sequence. 
	\end{proof}
\end{lemma}

To solve the optimization problem in \eqref{eq:N}, we propose the following construction for $\boldsymbol{N}$. 
\begin{enumerate}
	\item Input $N$, $M$.
	\item Let $N^*_{M-1}=2$, $N^*_0=N$, $m=M-1$
	\item If $m=1$, STOP; else continue.\label{step:it}
	\item $2\lfloor{(N^*_0-N^*_{m})/(2m)}\rfloor=\delta$
	\item If $\delta\geq N^*_m$, $N^*_{m-1}=2N^*_m$,\\
	else $N^*_{m-1}=N^*_m+\delta$.
	\item $m\rightarrow m-1$, return to Step \ref{step:it}
\end{enumerate}

For our construction of the remaining number of players $\boldsymbol{N^*}$, we have the following result.

\begin{theorem}
	The resulting $\boldsymbol{N^*}=\{N^*_m\}_{m=0}^{M-1}$ is a minimizer for the objective function \eqref{eq:N}.
	\begin{proof}
		From \autoref{thm:noninc}, we observe that the total amount of path change in the number of eliminated players is equal to the difference between the first number of eliminations and the last. Since at least $2$ players need to be eliminated and we have $N_{M-1}=2$, the last number of eliminated people is $2$. Thus, the problem is equivalent to minimizing the first number of eliminations. We can only achieve this by equally distributing the number of eliminations across rounds as much as possible, which concludes the proof.
	\end{proof}
\end{theorem}

\begin{example}
	The elimination numbers are found as follows.
	For example, given $N=134$ and $M=15$, we have
	\begin{align}
		&N_0=134,N_1=122,N_2=110,N_3=98,N_4=86,\\
		&N_5=76,N_6=66,N_7=56,N_8=46,N_{9}=36,\\
		&N_{10}=26,N_{11}=16,N_{12}=8,N_{13}=4,N_{14}=2
	\end{align}
\end{example}

\section{Discussions}

\begin{remark}
	Since at least two players are eliminated at each round, the same match never happens in consecutive rounds.
\end{remark}

Our design is build upon a seeded tournament setting. If we lack such a seed ranking there are a number of things that can be done.

\begin{itemize}
	\item One such example is a preliminary tournament with $\log_2(N)$ rounds to determine an initial ranking.
	
	\item Or we can consider a random initial seed, which will not be any worse than other tournament designs.
	
	\item Or we can modify the elimination and re-ranking rule accordingly for initial random seeds. In such scenarios, instead of decreasing the initial path change of $\boldsymbol{b}$ by $2$, we can consider a nonincreasing decrease where in the beginning stages, we can fully rank them. In such a scenario, eliminating fewer players at the beginning may be more meaningful.  	
\end{itemize}

\begin{remark}
	Moreover, the eliminated players can continue to play against each other in a separate league similar to our initial design to determine the full ranking instead of just the champion.
\end{remark}

\begin{remark}
	Furthermore, our tournament can be implemented in a connected manner by creating a losers bracket. We can design a promotion and relegation structure. Our near linear number of player elimination design will concern the net change in the players, i.e., the relegated players minus the promoted players.
\end{remark}

\begin{remark}
	Since the eliminations are only from the set of players that lost that round, most of the players face an elimination risk, which increases the stakes.
\end{remark}

\begin{remark}
	Our design is flexible in the sense that it can be adapted to any number of players $N$ as long as it is even and any number of matches $M$ as long as $\log_2 N\leq M\leq N/2$. If $N$ is odd, we can eliminate an odd number of players at the first round. If $M<\log_2 N$, the winner can be decided from the top-ranked remaining players. If $M>N/2$, the extra matches can be used for initial seeding.
\end{remark}

\bibliographystyle{IEEEtran}
\bibliography{double_bib}

\begin{thebibliography}{10}
\providecommand{\url}[1]{#1}
\csname url@samestyle\endcsname
\providecommand{\newblock}{\relax}
\providecommand{\bibinfo}[2]{#2}
\providecommand{\BIBentrySTDinterwordspacing}{\spaceskip=0pt\relax}
\providecommand{\BIBentryALTinterwordstretchfactor}{4}
\providecommand{\BIBentryALTinterwordspacing}{\spaceskip=\fontdimen2\font plus
\BIBentryALTinterwordstretchfactor\fontdimen3\font minus
  \fontdimen4\font\relax}
\providecommand{\BIBforeignlanguage}[2]{{%
\expandafter\ifx\csname l@#1\endcsname\relax
\typeout{** WARNING: IEEEtran.bst: No hyphenation pattern has been}%
\typeout{** loaded for the language `#1'. Using the pattern for}%
\typeout{** the default language instead.}%
\else
\language=\csname l@#1\endcsname
\fi
#2}}
\providecommand{\BIBdecl}{\relax}
\BIBdecl

\bibitem{sziklai2021efficacy}
B.~R. Sziklai, P.~Bir{\'o}, and L.~Csat{\'o}, ``The efficacy of tournament
  designs,'' \emph{arXiv preprint arXiv:2103.06023}, 2021.

\bibitem{lazear1981rank}
E.~P. Lazear and S.~Rosen, ``Rank-order tournaments as optimum labor
  contracts,'' \emph{Journal of political Economy}, vol.~89, no.~5, pp.
  841--864, 1981.

\bibitem{rosen1986prizes}
S.~Rosen, ``Prizes and incentives in elimination tournaments,'' \emph{The
  American Economic Review}, pp. 701--715, 1986.

\bibitem{taylor1995digging}
C.~R. Taylor, ``Digging for golden carrots: An analysis of research
  tournaments,'' \emph{The American Economic Review}, pp. 872--890, 1995.

\bibitem{prendergast1999provision}
C.~Prendergast, ``The provision of incentives in firms,'' \emph{Journal of
  economic literature}, vol.~37, no.~1, pp. 7--63, 1999.

\bibitem{szymanski2003economic}
S.~Szymanski, ``The economic design of sporting contests,'' \emph{Journal of
  economic literature}, vol.~41, no.~4, pp. 1137--1187, 2003.

\bibitem{orrison2004multiperson}
A.~Orrison, A.~Schotter, and K.~Weigelt, ``Multiperson tournaments: An
  experimental examination,'' \emph{Management Science}, vol.~50, no.~2, pp.
  268--279, 2004.

\bibitem{brown2014selecting}
J.~Brown and D.~B. Minor, ``Selecting the best? spillover and shadows in
  elimination tournaments,'' \emph{Management Science}, vol.~60, no.~12, pp.
  3087--3102, 2014.

\bibitem{bimpikis2019designing}
K.~Bimpikis, S.~Ehsani, and M.~Mostagir, ``Designing dynamic contests,''
  \emph{Operations Research}, vol.~67, no.~2, pp. 339--356, 2019.

\bibitem{klumpp2006primaries}
T.~Klumpp and M.~K. Polborn, ``Primaries and the new hampshire effect,''
  \emph{Journal of Public Economics}, vol.~90, no. 6-7, pp. 1073--1114, 2006.

\bibitem{ginsburgh2003expert}
V.~A. Ginsburgh and J.~C. Van~Ours, ``Expert opinion and compensation: Evidence
  from a musical competition,'' \emph{American Economic Review}, vol.~93,
  no.~1, pp. 289--296, 2003.

\bibitem{hou2021optimal}
T.~Hou and W.~Zhang, ``Optimal two-stage elimination contests for
  crowdsourcing,'' \emph{Transportation Research Part E: Logistics and
  Transportation Review}, vol. 145, p. 102156, 2021.

\bibitem{palacios2009field}
I.~Palacios-Huerta and O.~Volij, ``Field centipedes,'' \emph{American Economic
  Review}, vol.~99, no.~4, pp. 1619--35, 2009.

\bibitem{harris1987racing}
C.~Harris and J.~Vickers, ``Racing with uncertainty,'' \emph{The Review of
  Economic Studies}, vol.~54, no.~1, pp. 1--21, 1987.

\bibitem{yucesan2013efficient}
E.~Y{\"u}cesan, ``An efficient ranking and selection approach to boost the
  effectiveness of innovation contests,'' \emph{Iie Transactions}, vol.~45,
  no.~7, pp. 751--762, 2013.

\bibitem{ales2017optimal}
L.~Ales, S.-H. Cho, and E.~K{\"o}rpeo{\u{g}}lu, ``Optimal award scheme in
  innovation tournaments,'' \emph{Operations Research}, vol.~65, no.~3, pp.
  693--702, 2017.

\bibitem{hartigan1968inference}
J.~Hartigan, ``Inference from a knockout tournament,'' \emph{The Annals of
  Mathematical Statistics}, pp. 583--592, 1968.

\bibitem{israel1981stronger}
R.~B. Israel, ``Stronger players need not win more knockout tournaments,''
  \emph{Journal of the American Statistical Association}, vol.~76, no. 376, pp.
  950--951, 1981.

\bibitem{hwang1982new}
F.-K. Hwang, ``New concepts in seeding knockout tournaments,'' \emph{The
  American Mathematical Monthly}, vol.~89, no.~4, pp. 235--239, 1982.

\bibitem{horen1985comparing}
J.~Horen and R.~Riezman, ``Comparing draws for single elimination
  tournaments,'' \emph{Operations Research}, vol.~33, no.~2, pp. 249--262,
  1985.

\bibitem{knuth1987random}
D.~Knuth and O.~Lossers, ``A random knockout tournament (de knuth),''
  \emph{SIAM Review}, vol.~29, no.~1, pp. 127--129, 1987.

\bibitem{chen1988stronger}
R.~Chen and F.~Hwang, ``Stronger players win more balanced knockout
  tournaments,'' \emph{Graphs and Combinatorics}, vol.~4, no.~1, pp. 95--99,
  1988.

\bibitem{edwards1998non}
C.~T. Edwards, ``Non-parametric procedure for knockout tournaments,''
  \emph{Journal of Applied Statistics}, vol.~25, no.~3, pp. 375--385, 1998.

\bibitem{schwenk2000correct}
A.~J. Schwenk, ``What is the correct way to seed a knockout tournament?''
  \emph{The American Mathematical Monthly}, vol. 107, no.~2, pp. 140--150,
  2000.

\bibitem{glickman2008bayesian}
M.~E. Glickman, ``Bayesian locally optimal design of knockout tournaments,''
  \emph{Journal of Statistical Planning and Inference}, vol. 138, no.~7, pp.
  2117--2127, 2008.

\bibitem{vu2011fair}
T.~Vu and Y.~Shoham, ``Fair seeding in knockout tournaments,'' \emph{ACM
  Transactions on Intelligent Systems and Technology (TIST)}, vol.~3, no.~1,
  pp. 1--17, 2011.

\bibitem{groh2012optimal}
C.~Groh, B.~Moldovanu, A.~Sela, and U.~Sunde, ``Optimal seedings in elimination
  tournaments,'' \emph{Economic Theory}, vol.~49, no.~1, pp. 59--80, 2012.

\bibitem{prince2013designing}
M.~Prince, J.~Cole~Smith, and J.~Geunes, ``Designing fair 8-and 16-team
  knockout tournaments,'' \emph{IMA Journal of Management Mathematics},
  vol.~24, no.~3, pp. 321--336, 2013.

\bibitem{krakel2014optimal}
M.~Kr{\"a}kel, ``Optimal seedings in elimination tournaments revisited,''
  \emph{Economic Theory Bulletin}, vol.~2, no.~1, pp. 77--91, 2014.

\bibitem{hennessy2016bayesian}
J.~Hennessy and M.~Glickman, ``Bayesian optimal design of fixed knockout
  tournament brackets,'' \emph{Journal of Quantitative Analysis in Sports},
  vol.~12, no.~1, pp. 1--15, 2016.

\bibitem{karpov2016new}
A.~Karpov, ``A new knockout tournament seeding method and its axiomatic
  justification,'' \emph{Operations Research Letters}, vol.~44, no.~6, pp.
  706--711, 2016.

\bibitem{adler2017random}
I.~Adler, Y.~Cao, R.~Karp, E.~A. Pek{\"o}z, and S.~M. Ross, ``Random knockout
  tournaments,'' \emph{Operations Research}, vol.~65, no.~6, pp. 1589--1596,
  2017.

\bibitem{dagaev2018competitive}
D.~Dagaev and A.~Suzdaltsev, ``Competitive intensity and quality maximizing
  seedings in knock-out tournaments,'' \emph{Journal of Combinatorial
  Optimization}, vol.~35, no.~1, pp. 170--188, 2018.

\bibitem{karpov2018generalized}
A.~Karpov, ``Generalized knockout tournament seedings,'' \emph{International
  Journal of Computer Science in Sport}, vol.~17, no.~2, pp. 113--127, 2018.

\bibitem{arlegi2020fair}
R.~Arlegi and D.~Dimitrov, ``Fair elimination-type competitions,''
  \emph{European Journal of Operational Research}, vol. 287, no.~2, pp.
  528--535, 2020.

\bibitem{kulhanek2020surprises}
T.~Kulhanek and V.~Ponomarenko, ``Surprises in knockout tournaments,''
  \emph{Mathematics Magazine}, vol.~93, no.~3, pp. 193--199, 2020.

\bibitem{arlegi2022can}
R.~Arlegi, ``How can an elimination tournament favor a weaker player?''
  \emph{International Transactions in Operational Research}, vol.~29, no.~4,
  pp. 2250--2262, 2022.

\bibitem{harary1966theory}
F.~Harary and L.~Moser, ``The theory of round robin tournaments,'' \emph{The
  American Mathematical Monthly}, vol.~73, no.~3, pp. 231--246, 1966.

\bibitem{rubinstein1980ranking}
A.~Rubinstein, ``Ranking the participants in a tournament,'' \emph{SIAM Journal
  on Applied Mathematics}, vol.~38, no.~1, pp. 108--111, 1980.

\bibitem{gokcesu2017online}
K.~Gokcesu and S.~S. Kozat, ``Online anomaly detection with minimax optimal
  density estimation in nonstationary environments,'' \emph{IEEE Transactions
  on Signal Processing}, vol.~66, no.~5, pp. 1213--1227, 2017.

\bibitem{gokcesu2018sequential}
K.~Gokcesu, M.~M. Neyshabouri, H.~Gokcesu, and S.~S. Kozat, ``Sequential
  outlier detection based on incremental decision trees,'' \emph{IEEE
  Transactions on Signal Processing}, vol.~67, no.~4, pp. 993--1005, 2018.

\bibitem{cea2020analytics}
S.~Cea, G.~Dur{\'a}n, M.~Guajardo, D.~Saur{\'e}, J.~Siebert, and G.~Zamorano,
  ``An analytics approach to the fifa ranking procedure and the world cup final
  draw,'' \emph{Annals of Operations Research}, vol. 286, no.~1, pp. 119--146,
  2020.

\bibitem{goossens2012soccer}
D.~R. Goossens and F.~C. Spieksma, ``Soccer schedules in europe: an overview,''
  \emph{Journal of scheduling}, vol.~15, no.~5, pp. 641--651, 2012.

\bibitem{guyon2015rethinking}
J.~Guyon, ``Rethinking the fifa world cup™ final draw,'' \emph{Journal of
  Quantitative Analysis in Sports}, vol.~11, no.~3, pp. 169--182, 2015.

\bibitem{guyon2018fairer}
------, ``What a fairer 24 team uefa euro could look like,'' \emph{Journal of
  Sports Analytics}, vol.~4, no.~4, pp. 297--317, 2018.

\bibitem{guyon2020risk}
------, ``Risk of collusion: Will groups of 3 ruin the fifa world cup?''
  \emph{Journal of Sports Analytics}, vol.~6, no.~4, pp. 259--279, 2020.

\bibitem{kendall2010scheduling}
G.~Kendall, S.~Knust, C.~C. Ribeiro, and S.~Urrutia, ``Scheduling in sports: An
  annotated bibliography,'' \emph{Computers \& Operations Research}, vol.~37,
  no.~1, pp. 1--19, 2010.

\bibitem{laliena2019fair}
P.~Laliena and F.~J. L{\'o}pez, ``Fair draws for group rounds in sport
  tournaments,'' \emph{International Transactions in Operational Research},
  vol.~26, no.~2, pp. 439--457, 2019.

\bibitem{van2020handling}
D.~Van~Bulck and D.~Goossens, ``Handling fairness issues in time-relaxed
  tournaments with availability constraints,'' \emph{Computers \& Operations
  Research}, vol. 115, p. 104856, 2020.

\bibitem{wright2014or}
M.~Wright, ``Or analysis of sporting rules--a survey,'' \emph{European Journal
  of Operational Research}, vol. 232, no.~1, pp. 1--8, 2014.

\bibitem{chater2021fixing}
M.~Chater, L.~Arrondel, J.-P. Gayant, and J.-F. Laslier, ``Fixing match-fixing:
  Optimal schedules to promote competitiveness,'' \emph{European Journal of
  Operational Research}, vol. 294, no.~2, pp. 673--683, 2021.

\bibitem{csato2020incentive}
L.~Csat{\'o}, ``The incentive (in) compatibility of group-based qualification
  systems,'' \emph{International Journal of General Systems}, vol.~49, no.~4,
  pp. 374--399, 2020.

\bibitem{csato2021tournament}
------, \emph{Tournament design: How operations research can improve sports
  rules}.\hskip 1em plus 0.5em minus 0.4em\relax Springer Nature, 2021.

\bibitem{csato2022quantifying}
------, ``Quantifying incentive (in) compatibility: A case study from sports,''
  \emph{European Journal of Operational Research}, 2022.

\bibitem{dagaev2018winning}
D.~Dagaev and K.~Sonin, ``Winning by losing: Incentive incompatibility in
  multiple qualifiers,'' \emph{Journal of Sports Economics}, vol.~19, no.~8,
  pp. 1122--1146, 2018.

\bibitem{pauly2014can}
M.~Pauly, ``Can strategizing in round-robin subtournaments be avoided?''
  \emph{Social Choice and Welfare}, vol.~43, no.~1, pp. 29--46, 2014.

\bibitem{preston2003cheating}
I.~Preston and S.~Szymanski, ``Cheating in contests,'' \emph{Oxford review of
  economic policy}, vol.~19, no.~4, pp. 612--624, 2003.

\bibitem{vong2017strategic}
A.~I. Vong, ``Strategic manipulation in tournament games,'' \emph{Games and
  Economic Behavior}, vol. 102, pp. 562--567, 2017.

\bibitem{krumer2020testing}
A.~Krumer, ``Testing the effect of kick-off time in the uefa europa league,''
  \emph{European Sport Management Quarterly}, vol.~20, no.~2, pp. 225--238,
  2020.

\bibitem{atan2018minimization}
T.~Atan and B.~{\c{C}}avdaro{\u{g}}lu, ``Minimization of rest mismatches in
  round robin tournaments,'' \emph{Computers \& Operations Research}, vol.~99,
  pp. 78--89, 2018.

\bibitem{bergantinos2020sharing}
G.~Bergantinos and J.~D. Moreno-Ternero, ``Sharing the revenues from
  broadcasting sport events,'' \emph{Management Science}, vol.~66, no.~6, pp.
  2417--2431, 2020.

\bibitem{petroczy2021revenue}
D.~G. Petr{\'o}czy and L.~Csat{\'o}, ``Revenue allocation in formula one: a
  pairwise comparison approach,'' \emph{International Journal of General
  Systems}, vol.~50, no.~3, pp. 243--261, 2021.

\bibitem{gokcesu2021efficient}
K.~Gokcesu and H.~Gokcesu, ``Efficient locally optimal number set partitioning
  for scheduling, allocation and fair selection,'' \emph{arXiv preprint
  arXiv:2109.04809}, 2021.

\bibitem{gokcesu2021quadratic}
------, ``A quadratic time locally optimal algorithm for np-hard equal
  cardinality partition optimization,'' \emph{arXiv preprint arXiv:2109.07882},
  2021.

\bibitem{gokcesu2022linearithmic}
------, ``A linearithmic time locally optimal algorithm for the multiway number
  partition optimization,'' \emph{arXiv preprint arXiv:2203.05618}, 2022.

\bibitem{gokcesu2021optimally}
------, ``Optimally efficient sequential calibration of binary classifiers to
  minimize classification error,'' \emph{arXiv preprint arXiv:2108.08780},
  2021.

\bibitem{gokcesu2021anytime}
------, ``Efficient, anytime algorithms for calibration with isotonic
  regression under strictly convex losses,'' \emph{arXiv preprint
  arXiv:2111.00468}, 2021.

\bibitem{gokcesu2021generalized}
------, ``Generalized huber loss for robust learning and its efficient
  minimization for a robust statistics,'' \emph{arXiv preprint
  arXiv:2108.12627}, 2021.

\bibitem{gokcesu2022nonconvex}
------, ``Nonconvex extension of generalized huber loss for robust learning and
  pseudo-mode statistics,'' \emph{arXiv preprint arXiv:2202.11141}, 2022.

\bibitem{gokcesu2021regret}
------, ``Regret analysis of global optimization in univariate functions with
  lipschitz derivatives,'' \emph{arXiv preprint arXiv:2108.10859}, 2021.

\bibitem{gokcesu2022low}
------, ``Low regret binary sampling method for efficient global optimization
  of univariate functions,'' \emph{arXiv preprint arXiv:2201.07164}, 2022.

\bibitem{gokcesu2022natural}
------, ``Natural hierarchical cluster analysis by nearest neighbors with
  near-linear time complexity,'' \emph{arXiv preprint arXiv:2203.08027}, 2022.

\bibitem{gokcesu2021nonparametric}
------, ``Nonparametric extrema analysis in time series for envelope
  extraction, peak detection and clustering,'' \emph{arXiv preprint
  arXiv:2109.02082}, 2021.

\bibitem{david1959tournaments}
H.~A. David, ``Tournaments and paired comparisons,'' \emph{Biometrika},
  vol.~46, no. 1/2, pp. 139--149, 1959.

\bibitem{glenn1960comparison}
W.~Glenn, ``A comparison of the effectiveness of tournaments,''
  \emph{Biometrika}, vol.~47, no. 3/4, pp. 253--262, 1960.

\bibitem{searls1963probability}
D.~T. Searls, ``On the probability of winning with different tournament
  procedures,'' \emph{Journal of the American Statistical Association},
  vol.~58, no. 304, pp. 1064--1081, 1963.

\bibitem{appleton1995may}
D.~R. Appleton, ``May the best man win?'' \emph{Journal of the Royal
  Statistical Society: Series D (The Statistician)}, vol.~44, no.~4, pp.
  529--538, 1995.

\bibitem{mcgarry1997efficacy}
T.~McGarry and R.~Schutz, ``Efficacy of traditional sport tournament
  structures,'' \emph{Journal of the Operational Research Society}, vol.~48,
  no.~1, pp. 65--74, 1997.

\bibitem{mendoncca2000comparing}
D.~Mendon{\c{c}}a and M.~Raghavachari, ``Comparing the efficacy of ranking
  methods for multiple round-robin tournaments,'' \emph{European Journal of
  Operational Research}, vol. 123, no.~3, pp. 593--605, 2000.

\bibitem{marchand2002comparison}
{\'E}.~Marchand, ``On the comparison between standard and random knockout
  tournaments,'' \emph{Journal of the Royal Statistical Society: Series D (The
  Statistician)}, vol.~51, no.~2, pp. 169--178, 2002.

\bibitem{ryvkin2008predictive}
D.~Ryvkin and A.~Ortmann, ``The predictive power of three prominent tournament
  formats,'' \emph{Management Science}, vol.~54, no.~3, pp. 492--504, 2008.

\bibitem{ryvkin2010selection}
D.~Ryvkin, ``The selection efficiency of tournaments,'' \emph{European Journal
  of Operational Research}, vol. 206, no.~3, pp. 667--675, 2010.

\bibitem{scarf2009numerical}
P.~Scarf, M.~M. Yusof, and M.~Bilbao, ``A numerical study of designs for
  sporting contests,'' \emph{European Journal of Operational Research}, vol.
  198, no.~1, pp. 190--198, 2009.

\bibitem{scarf2011numerical}
P.~A. Scarf and M.~M. Yusof, ``A numerical study of tournament structure and
  seeding policy for the soccer world cup finals,'' \emph{Statistica
  Neerlandica}, vol.~65, no.~1, pp. 43--57, 2011.

\bibitem{goossens2012comparing}
D.~R. Goossens, J.~Beli{\"e}n, and F.~C. Spieksma, ``Comparing league formats
  with respect to match importance in belgian football,'' \emph{Annals of
  Operations Research}, vol. 194, no.~1, pp. 223--240, 2012.

\bibitem{annis2006comparison}
D.~H. Annis and S.~S. Wu, ``A comparison of potential playoff systems for ncaa
  ia football,'' \emph{The American Statistician}, vol.~60, no.~2, pp.
  151--157, 2006.

\bibitem{lasek2018efficacy}
J.~Lasek and M.~Gagolewski, ``The efficacy of league formats in ranking
  teams,'' \emph{Statistical Modelling}, vol.~18, no. 5-6, pp. 411--435, 2018.

\bibitem{csato2021simulation}
L.~Csat{\'o}, ``A simulation comparison of tournament designs for the world
  men's handball championships,'' \emph{International Transactions in
  Operational Research}, vol.~28, no.~5, pp. 2377--2401, 2021.

\bibitem{engist2021effect}
O.~Engist, E.~Merkus, and F.~Schafmeister, ``The effect of seeding on
  tournament outcomes: Evidence from a regression-discontinuity design,''
  \emph{Journal of Sports Economics}, vol.~22, no.~1, pp. 115--136, 2021.

\end{thebibliography}
\end{document}